# Automatically Evaluating Balance: A Machine Learning Approach

Tian Bao, Brooke N. Klatt, Susan L. Whitney, Kathleen H. Sienko, Jenna Wiens

*Abstract*—Compared to in-clinic balance training, in-home training is not as effective. This is, in part, due to the lack of feedback from physical therapists (PTs). Here, we analyze the feasibility of using trunk sway data and machine learning (ML) techniques to automatically evaluate balance, providing accurate assessments outside of the clinic. We recruited sixteen participants to perform standing balance exercises. For each exercise, we recorded trunk sway data and had a PT rate balance performance on a scale of 1 to 5. The rating scale was adapted from the Functional Independence Measure. From the trunk sway data, we extracted a 61-dimensional feature vector representing performance of each exercise. Given these labeled data, we trained a multi-class support vector machine (SVM) to map trunk sway features to PT ratings. Evaluated in a leave-one-participant-out scheme, the model achieved a classification accuracy of 82%. Compared to participant self-assessment ratings, the SVM outputs were significantly closer to PT ratings. The results of this pilot study suggest that in the absence of PTs, ML techniques can provide accurate assessments during standing balance exercises. Such automated assessments could reduce PT consultation time and increase user compliance outside of the clinic.

## I. INTRODUCTION

FALLs caused by balance impairments lead to loss of mobility, anxiety, and reduced quality of life [1]–[6]. Balance training programs can improve balance in people with balance deficits (e.g., vestibular disorders, Parkinson's disease, stroke) [7]–[11]. Physical therapists (PTs) use more than 200 static standing, dynamic standing and gait balance exercises as part of rehabilitation programs, with levels of difficulty ranging from easy (1) to hard (5) [12], [13]. Balance exercises are generated by varying the stance, visual inputs, standing surface and head motion [12]. In clinics, PTs select and customize balance exercises for each individual [12]. At home, individuals typically perform balance exercises based on either a PT's instructions during in-clinic training or paper instructions [14]. However, due to the lack of supervision and consequent loss of motivation, in-home training is not as effective as in-clinic training, and therefore does not lead to the same improvements in balance-related function outcomes [15]–[17].

To improve the effectiveness of in-home balance training, researchers have introduced remote supervision (e.g., telerehabilitation), sensory augmentation (e.g., video game–based balance training), and/or semi-supervision (e.g., wearable devices with periodic expert input) [18], [19]. Telerehabilitation delivers remote rehabilitation services, including assessments and interventions, via telecommunication networks [18]. Cikajlo et al. showed that for people with stroke, telerehabilitation with virtual reality tasks led to balance improvements similar to those associated with conventional in-clinic balance training [20]. Telerehabilitation, however, generally requires remote interactions between the expert and user for the duration of the training session via video conference that can result in longer PT consultation times [21]. Video game–based in-home balance training has been shown to improve clinical measures after a minimum of five weeks of training [22]–[25]. Kinetic or kinematic data are collected using balance platforms (e.g., Wii Fit balance board) or video cameras with depth sensors (e.g., Kinect) and a display screen to provide visual cues of balance. Although visual cues are effective for a large number of rehabilitation exercises, there are a subset of rehabilitation exercises that are performed with eyes closed and/or head movements. Additionally, balance platforms such as the Wii Fit balance board, constrain the types of stance positions that can be performed. Recently, Bao et al. demonstrated improvements in clinical balance outcomes using a semi-supervised balance training protocol for older adults in their homes; both the experimental group that received vibrotactile sensory augmentation during balance training via a smartphone balance trainer (i.e., vibrotactile sensory augmentation device) and a control group that performed the exercises unaided showed significant improvements in clinical outcomes [13]. PTs remotely determined exercise progression by reviewing participant data on a weekly basis. PTs prescribed exercises for the following week based on the number of step-outs from the correct position and participant self-assessments on a 1-5 scale. However, participant self-assessments may not align with PT assessments [26], [27]. This misalignment can affect the appropriateness of selected exercises and limit the overall effectiveness of a balance training program.

Appropriate progression of balance exercises within a balance-training regimen is critical to achieving improvements in balance for both clinic- and home-based settings [12]. The extent to which the human (here, the expert/PT) is "in the loop"

Tian Bao is with the Department of Mechanical Engineering, University of Michigan, Ann Arbor, MI 48109 USA (email: baotian@umich.edu).

Brooke N. Klatt and Susan L. Whitney are with the Department of Physical Therapy and Otolaryngology, School of Health and Rehabilitation Sciences, University of Pittsburgh, Pittsburgh, PA 15260 USA (e-mail: bnk12@pitt.edu and whitney@pitt.edu).

Kathleen H. Sienko is with the Department of Mechanical Engineering, University of Michigan, Ann Arbor, MI 48109 USA (e-mail: sienko@umich.edu).

Jenna Wiens is with the Department of Computer Science and Engineering, University of Michigan, Ann Arbor, MI 48109 USA (*corresponding author, e-mail: wiensj@umich.edu).

is diminished as one shifts from a telerehabilitation to a semi-supervised balance training program paradigm [28]. The use of machine learning (ML) techniques could augment this setting further supporting home-based balance rehabilitation training.

Recently, researchers have demonstrated the utility of ML in supporting clinical assessments of gait based on body motion [29]–[35]. Using motion data from wearable sensors, researchers have successfully leveraged ML techniques to identify activity [29], mine gait patterns [30], and classify gait disorders [31]–[34]. For example, Begg and Owen successfully applied supervised learning techniques to automatically recognize young versus old gait types, based on data collected from a motion capture system, with 83.3% accuracy [31]. Similarly, LeMoyne et al. used features extracted from motion data measured by an inertial measurement unit (IMU) on the ankle joint to train a neural network to distinguish between older adults and people with Friedreich's ataxia [32].

Given the successful use of ML-based techniques for activity and gait applications to date, we applied ML techniques in this study to learn an accurate mapping from trunk sway collected by an IMU to a PT's assessment ratings of balance performance and assessed the feasibility of an ML-based approach for automatically and accurately evaluating balance exercise performance.

## II. METHODS

### A. Data

Sixteen participants (68.2±8.0 yrs, five males, 11 females) with balance deficits or balance concerns (participant demographics are shown in Table I) performed balance exercises during 18 sessions over the course of six weeks. Participants with balance deficits including a diagnosed vestibular disorder or peripheral neuropathy were recruited by PTs and via flyers at the university medical center. Participants were excluded if they had confounding neurologic or neuromuscular disorders; known pregnancy; recent lower extremity fractures/severe sprains (within the last six months); previous lower extremity joint replacement; incapacitating back or lower extremity pain; the inability to stand for three minutes without rest; a body habitus that exceeded the dimensions of the measuring equipment (waist circumference >50 inches); or a Montreal Cognitive Assessment score of less than 26 points [36]. All participants gave written informed consent, and the study was conducted in accordance with the Declaration of Helsinki. The study protocol was reviewed and approved by the University of Pittsburgh Institutional Review Board (PRO13020399). In each session, participants performed multiple trials of two standing exercises. Exercises were selected by a PT from a set of 60 standard standing exercises for balance rehabilitation [12]. The PT used clinical judgment to select exercises with a moderate difficulty level. The exercises were generated by varying the visual (open/closed), stance (feet apart/feet together/semi-tandem Romberg/tandem Romberg/single leg stance), head motion (none/pitch/yaw), and support surface (firm/foam) conditions [12], [13]. Each participant performed six trials of a given exercise, and each trial lasted for 30 seconds. If participants had to step out or needed help in maintaining balance, the trial was terminated and marked as a "step-out" trial. After each set of six trials, the PT and the participant who performed the exercise each provided one balance performance assessment rating across the six trials. The rating rubrics for both scales are shown in Table II. The text accompanying the numerical values of the two scales used in this study varied based on the end user (participant performing the balance task or expert assessing the balance task), but aimed to capture a similar level of performance. The participant and expert scales were adapted from Espy et al. [37] and the Functional Independence Measure [38], respectively. One scale that more closely resembled the expert scale was initially piloted for both the participants and the experts; however, the participants had difficulty using non-laymen phrasing, motivating the inclusion of a participant-specific scale that better reflected language they would use to describe their performance. Both scales were intentionally designed to have 5 points based on previously published scales [37], [38].

During training, participants donned a wearable IMU (MTx, Xsens Technologies B.V.) aligned with the L4/L5 spinal segment level on their back. The IMU comprised accelerometers and gyroscopes to estimate trunk sway relative to the gravitational vector in both the pitch and roll directions in real-time [39]. Fig. 1 shows an example of trunk sway in the pitch and roll directions recorded during an exercise trial.

TABLE I
DEMOGRAPHIC INFORMATION FOR PARTICIPANTS

| ID | Age | Gender | Diagnosis | MOCA Score |
|----|-----|--------|-----------|------------|
| 1 | 78 | Female | Bilateral Vestibular Disorder | 26 |
| 2 | 47 | Female | Unilateral Vestibular Disorder | 26 |
| 3 | 67 | Female | Unilateral Vestibular Disorder | 26 |
| 4 | 68 | Female | Unilateral Vestibular Disorder | 29 |
| 5 | 70 | Female | Peripheral Neuropathy | 28 |
| 6 | 79 | Female | Balance Concerns | 26 |
| 7 | 69 | Male | Peripheral Neuropathy | 28 |
| 8 | 65 | Male | Balance Concerns | 27 |
| 9 | 63 | Female | Unilateral Vestibular Disorder | 28 |
| 10 | 65 | Female | Peripheral Neuropathy | 30 |
| 11 | 61 | Male | Balance Concerns | 26 |
| 12 | 79 | Male | Unilateral Vestibular Disorder | 30 |
| 13 | 74 | Female | Bilateral Vestibular Disorder | 26 |
| 14 | 69 | Male | Peripheral Neuropathy | 29 |
| 15 | 74 | Female | Unilateral Vestibular Disorder | 30 |
| 16 | 63 | Female | Unilateral Vestibular Disorder | 27 |

TABLE II
RATING RUBRICS FOR THE PHYSICAL THERAPIST (ADAPTED FROM THE FUNCTIONAL INDEPENDENCE MEASURE [38]) AND PARTICIPANTS (ADAPTED FROM ESPY ET AL. [37])

| Ratings | Description for physical therapist | Description for participants |
|---------|-----------------------------------|------------------------------|
| 1 | Independent with no sway | I feel completely steady |
| 2 | Supervision with minimal sway | I feel a little unsteady or off-balance |
| 3 | Close supervision with moderate sway | I feel somewhat unsteady, or like I may lose my balance |
| 4 | Requires physical assistance or positive stepping strategy after 15 seconds | I feel very unsteady, or like I definitely will lose my balance |
| 5 | Unable to maintain position with assist or step out in the first 15 seconds of the exercise | I lost my balance |

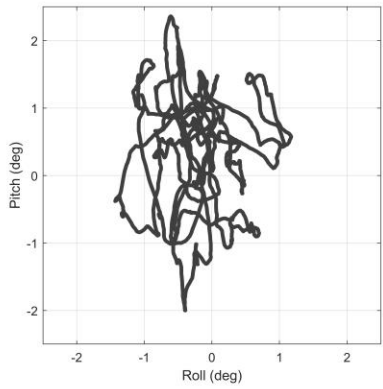

Fig. 1. An example of trunk sway recorded during an exercise trial (feed apart stance on foam surface with eyes closed). In this particular example, the participant demonstrates more sway (i.e., variation) in the pitch direction relative to the roll direction.

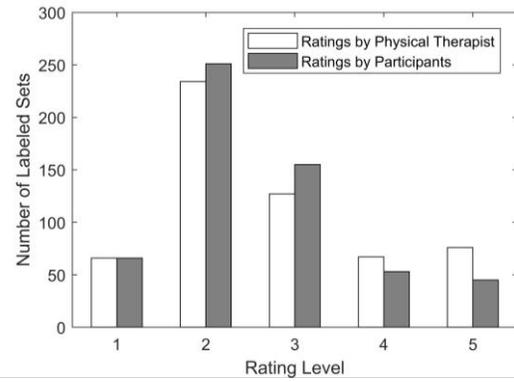

Fig. 2. Distribution of rating levels provided by the physical therapist and participants.

In total, 16 participants performed 576 exercises (16 participants x 18 sessions x 2 exercises). However, technical issues resulted in the loss of trunk sway data for three sessions. After discarding those three sessions, 570 labeled sets remained. Each labeled set contained trunk sway data, the PT's rating, and a participant's rating. Fig. 2 shows the number of sets included in the dataset as a function of the rating levels provided by the PT and participants.

### B. Data features

To characterize trunk motion we calculated ten common kinematic metrics (Fig. 3) for each of the six trials within each balance exercise set [40]–[47]. Since we only obtained one PT rating and one participant rating for each set of six trials, we summarized the kinematic metrics across the six trials within each balance exercise set by computing six statistical descriptors: 1) mean, 2) standard derivation, 3) minimum, 4) maximum, 5) median, and 6) trend (i.e., up: 1, flat: 0, or down: -1) for all ten metrics (Fig. 3). Trend was calculated based on the sign of the slope to the linear fit. If the magnitude of the slope was less than 0.005, the trend was considered flat. We added the *number of non-"step-out" trials* across the six trials as the last feature. This resulted in 61 continuous valued features (Fig. 3) for each set of trials. Since these features lay on different scales, we z-scaled all features to zero mean and unit standard deviation. All data preprocessing was completed using MATLAB (MathWorks, R2016b).

### C. Classification

Using the data above, we aimed to predict the PT's ratings based on trunk sway by learning a mapping from the 61-dimensional feature space to the PT's ratings. To learn this mapping, we trained a multi-class support vector machine (SVM) with a linear kernel [48]. We also considered other techniques (e.g., extreme gradient boosting (XGBoost) [49], SVM with non-linear kernels, support vector regression). However, these techniques did not yield statistically significantly different results from the linear SVM approach and lacked interpretability. To build the multi-class SVM, we learned a separate classifier for each pair of classes. Here, we used a one-vs-one framework since it is less sensitive to an imbalanced dataset compared to a one-vs-all framework. To account for the variation in frequency across classes, we used asymmetric costs [50]. The asymmetric costs were set to the inverse frequency of each class (i.e., rating level). Given the ten classifiers (e.g., 1 vs. 2, 1 vs. 3, etc.), test examples were then classified by applying each classifier in turn and taking the majority vote.

### D. Validation & Evaluation

We used the leave-one-participant-out method for both validation and evaluation. During the validation phase we tuned the hyperparameter and during the evaluation phase we evaluated model performance. The use of leave-one-participant-out eliminates bias due to inter-participant differences [51]. The details of this process are shown in Fig. 4. The data of each participant, in turn, were marked as the test dataset, and the remaining data were used for training and validation. Within each training dataset, we used leave-one-participant-out cross-validation to tune the hyperparameter. More specifically, we tuned the SVM cost parameter ($C$), sweeping from [1e-7,1e3]. We averaged the performance across cross-validation folds, and selected the $C$ that led to the best averaged F1 score. We used the averaged F1 score because of the imbalanced distribution of rating levels as demonstrated in Fig. 2. Given this optimal $C$, we trained a final classifier on the training data and then applied it to the test dataset. We repeated the process for each participant (i.e., repeated 16 times) to obtain the final overall classification accuracy and averaged F1 score.

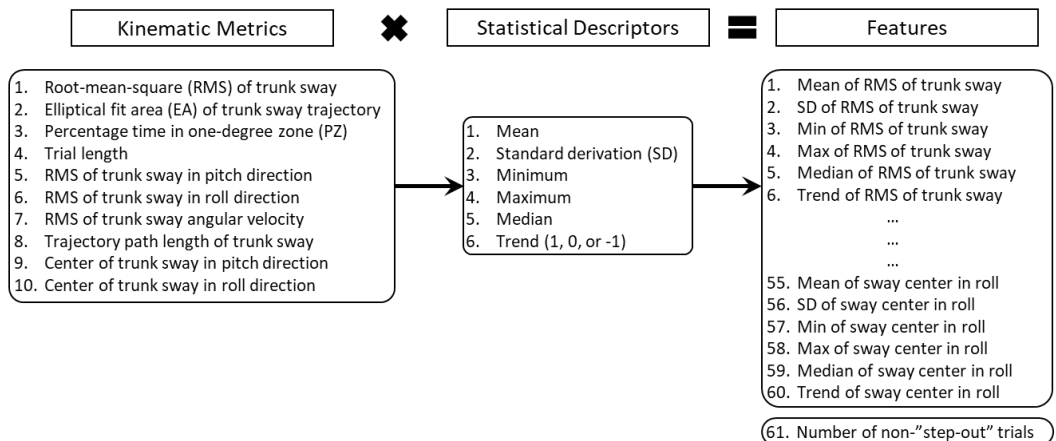

Fig. 3. Sixty-one features were calculated to summarize performance across each set of six trials based on ten commonly used kinematic metrics and six statistical descriptors.

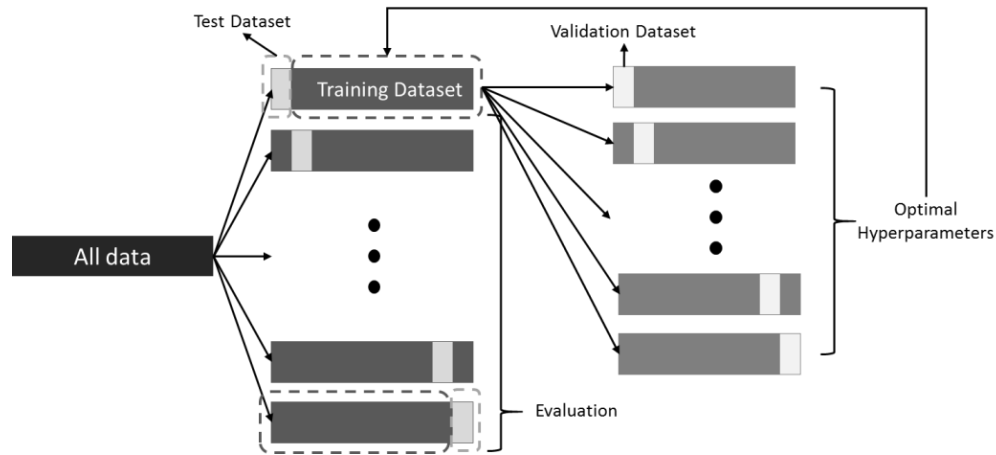

Fig. 4. Validation and evaluation of the classifier using the leave-one-participant-out cross validation. For visualization, only the first cross-validation is shown.

The averaged F1 score was obtained by averaging the F1 scores of all rating levels. We calculated the F1 score for each rating level using the precision and the recall at that rating level (Eqs 1, 2, and 3).

$$F1 \text{ score} = 2 \cdot \frac{precision \cdot recall}{precision + recall} \quad \text{Eq 1}$$

$$Precision = \frac{True\ Positive}{True\ Positive + False\ Positive} \quad \text{Eq 2}$$

$$Recall = \frac{True\ Positive}{True\ Positive + False\ Negative} \quad \text{Eq 3}$$

### E. Feature importance

After evaluating model performance, we used backward feature elimination to measure the relative importance of the features. Due to collinearity among the features, we cannot simply interpret the learned model coefficients as "importance." To ascertain the contribution each feature made (relative to the other features), we used backward feature elimination [52], [53]. This approach is commonly used as a dimensionality reduction method and can highlight the effects of including or removing a feature in the context of the others. To avoid overfitting, we used a leave-one-participant-out setup similar to the one shown in Fig. 4. For each training dataset, features were eliminated one by one based on model performance (i.e., averaged F1 score) until only a single feature remained. This gave us a measure of the relative importance of each feature, where the first feature to be eliminated had a relative importance of 61 (i.e., least important), and the last feature to be eliminated resulted in a relative importance of 1 (i.e., most important). This feature elimination procedure was repeated for each participant (i.e., 16 times). By averaging the relative importance of each feature (i.e., 16 relative importance), we obtained an overall ranking of features. To understand the relative importance of each metric (e.g., trial length) and each statistical descriptor (e.g., mean), we averaged the relative importance of features within each category (i.e., the 10 metrics and 6 statistical descriptors shown in Fig. 3).

### F. Additional classification task

Based on input from three PTs involved in this study, we considered an additional three-level classification task in which we grouped rating levels 1 and 2 together, and rating levels 3 and 4 together. In general, PTs identified exercises rated 1 or 2 as easy/safe, exercises rated 3 or 4 as moderate/suitable, and exercises rated 5 as difficult/unsafe.

### G. Statistical analysis

To compare the agreement between the SVM and PT assessments (i.e., classification accuracy) with the agreement between the PT and self-assessments, we used a paired two-tailed t-test with a significance level of 0.05. Similarly, when comparing the different machine learning techniques (mentioned earlier), we used a paired two-tailed t-test. All classification and statistical analyses were performed using R (r-project.org).

## III. RESULTS

The performance of the five-class and three-class classification models is shown in Table III. For the five-class classification task, the SVM predictions agreed with the PT's ratings with an accuracy of 64.3% and averaged F1 score of 0.64. The accuracy was 13.8 percentage points ($p<0.001$) better than the agreement achieved by the self-assessment ratings, while the averaged F1 score was 0.18 better than the agreement achieved by the self-assessment ratings. After grouping ratings into three-classes, the accuracy and averaged F1 score improved to 82.0% and 0.81, respectively. Moreover, the SVM still outperformed the participants' self-ratings (13.2 percentage points (*p<0.001*) for accuracy and 0.19 for the averaged F1 score).

TABLE III
RESULTS FOR THE FIVE-LEVEL AND THREE-LEVEL CLASSIFICATION MODELS. STANDARD DEVIATIONS ARE REPORTED AND THE SYMBOL * INDICATES STATISTICAL SIGNIFICANCE (P<0.05).

| Models | Five-class Classification | | Three-class Classification | |
|---|---|---|---|---|
| | Accuracy* | Averaged F1 Score | Accuracy* | Averaged F1 Score |
| Participant | 50.6±08.1% | 0.46±0.16 | 68.8±9.8% | 0.62±0.15 |
| SVM | **64.3±11.3%** | **0.64±0.11** | **82.0±8.4%** | **0.81±0.07** |

The confusion matrices for the self-assessments and five-class SVM classifier predictions are shown in Table IV. The calculated F1 score, precision, and recall for each rating level are also shown with the confusion matrices. The SVM predictions outperformed participants' self-assessments for all rating levels in terms of the F1 scores.

Table V shows a) the top ten features ranked by relative importance, b) all metrics ranked by relative importance, and c) all statistical descriptors ranked by relative importance obtained from the backward feature elimination method. The *number of non-"step-out" trials* and *min of RMS in roll* were the two most important features among all 61 features. When using the *number of step-out trials* (non-IMU feature) alone, the accuracy dropped to 45% and the average F1 score dropped to 0.5.

## IV. DISCUSSION

In this pilot study, we used an SVM-based approach to automatically assess balance performance based on trunk sway collected from an IMU sensor. To the best of our knowledge, this is the first study to develop and validate the use of ML techniques to automatically provide PT-like assessments of balance performance. On held-out data, compared to self-assessment ratings, the ratings generated by the SVM classifier were significantly closer to those of the PT.

In the five-class rating classification, the SVM outperformed the participants' self-assessments with respect to precision, recall, and F1-score at all rating levels with the exception of precision in rating level 1 and recall in rating level 2. Increased precision indicates a higher confidence when exercises are predicted at a particular rating level. Increased recall indicates that the model correctly captures a larger portion of the exercises at this rating level. Compared to self-assessments, the recall rate increased from 0.42 to 0.78 and the precision rate increased from 0.71 to 0.86 when using the SVM for rating level 5. This indicates that the SVM model was significantly better than participants at identifying difficult/unsafe exercises that could cause losses of balance. The recall for exercises rated as level 2 and precision for exercises rated as level 1 decreased slightly compared to the self-assessments. However, upon further inspection we noted that the algorithm re-classified many exercises rated level 2 as level 1. Further investigation is needed to understand the difference between rating levels 1 and 2. Such a distinction is particularly important, since when designing a balance training program, the inclusion of easy to perform exercises (i.e., exercises with low ratings (1,2)) should be minimized because they produce fewer benefits [54]. Exercises rated

TABLE IV
CONFUSION MATRIX FOR A) PARTICIPANTS' SELF-ASSESSMENTS VERSUS THE PHYSICAL THERAPIST'S ASSESSMENTS AND B) SVM PREDICTIONS VERSUS PHYSICAL THERAPIST'S ASSESSMENTS

**A**

| | | Self-assessments | | | | | | Recall |
|---|---|---|---|---|---|---|---|---|
| | | 1 | 2 | 3 | 4 | 5 | Sum | |
| PT assessments | 1 | 36 | 29 | 1 | 0 | 0 | 66 | 0.55 |
| | 2 | 27 | 154 | 51 | 2 | 0 | 234 | 0.66 |
| | 3 | 3 | 49 | 53 | 21 | 1 | 127 | 0.42 |
| | 4 | 0 | 15 | 26 | 14 | 12 | 67 | 0.21 |
| | 5 | 0 | 4 | 24 | 16 | 32 | 76 | 0.42 |
| | Sum | 66 | 251 | 155 | 53 | 45 | 570 | - |
| Precision | | 0.55 | 0.61 | 0.34 | 0.27 | 0.71 | - | Accuracy: 51% |
| F1 Score | | 0.55 | 0.64 | 0.38 | 0.23 | 0.53 | - | |

**B**

| | | SVM Predictions | | | | | | Recall |
|---|---|---|---|---|---|---|---|---|
| | | 1 | 2 | 3 | 4 | 5 | Sum | |
| PT assessments | 1 | 56 | 10 | 0 | 0 | 0 | 66 | 0.85 |
| | 2 | 53 | 142 | 33 | 6 | 0 | 234 | 0.61 |
| | 3 | 3 | 29 | 68 | 26 | 1 | 127 | 0.54 |
| | 4 | 1 | 3 | 12 | 42 | 9 | 67 | 0.63 |
| | 5 | 0 | 0 | 3 | 14 | 59 | 76 | 0.78 |
| | Sum | 113 | 184 | 116 | 88 | 69 | 570 | - |
| Precision | | 0.5 | 0.77 | 0.59 | 0.48 | 0.86 | - | Accuracy: 64% |
| F1 Score | | 0.63 | 0.68 | 0.56 | 0.54 | 0.81 | - | |

TABLE V
(A) THE TOP TEN FEATURES AND THEIR RELATIVE IMPORTANCE; (B) THE RELATIVE IMPORTANCE OF EACH METRIC; (C) THE RELATIVE IMPORTANCE OF EACH STATISTICAL DESCRIPTOR.

**A**

| Rank | Features | Importance Ranking |
|---|---|---|
| 1 | # of Non-"Step-out" Trials | 1.1 ± 0.5 |
| 2 | Min of RMS in Roll | 2.3 ± 0.7 |
| 3 | Trend of RMS | 7.9 ± 7.6 |
| 4 | Mean of Trial Length | 10.9 ± 11.3 |
| 5 | Trend of PZ | 12.2 ± 6.7 |
| 6 | Trend of Path Length | 12.8 ± 10.6 |
| 7 | Trend of Center of Sway in Pitch | 15.6 ± 5.3 |
| 8 | Trend of RMS in Roll | 17.3 ± 11.4 |
| 9 | Trend of EA | 18.7 ± 11.0 |
| 10 | Min of EA | 19.7 ± 12.9 |

**B**

| Rank | Metrics | Importance Ranking |
|---|---|---|
| 1 | RMS in Roll | 21.7±12.0 |
| 2 | PZ | 27.2±8.4 |
| 3 | Trail Length | 28.6±12.5 |
| 4 | Path Length | 31.0±11.5 |
| 5 | Center of Sway in Roll | 32.3±7.8 |
| 6 | EA | 33.5±11.9 |
| 7 | RMS in Pitch | 34.6±6.9 |
| 8 | Center of Sway in Pitch | 34.7±13.3 |
| 9 | RMS of trunk sway | 34.8±14.9 |
| 10 | RMS of Velocity | 36.6±6.8 |

**C**

| Rank | Statistical Descriptors | Importance Ranking |
|---|---|---|
| 1 | Trend | 19.4±7.7 |
| 2 | Min | 29.1±11.5 |
| 3 | Median | 31.8±6.8 |
| 4 | Standard Deviation | 33.6±8.4 |
| 5 | Max | 37.0±6.0 |
| 6 | Mean | 38.0±12.0 |

level 3 or level 4 (i.e., exercises most often recommended by PTs for repetitive training) showed both increased precision and recall.

Balance performance is often described by kinematic metrics including the RMS of trunk sway, PZ, EA of trunk sway, trial length, trajectory path length of trunk sway, and RMS of trunk sway angular velocity [40]–[47]. Here, we investigated the relative importance of different kinematic metrics and statistical descriptors to better understand how they relate to PT assessments. In this study, RMS of trunk sway, PZ, trial length, and path length were ranked as the top four indicators. Notably, RMS of trunk sway in the roll direction was more important than overall RMS and RMS in pitch direction, which suggests that RMS in the roll direction is more closely associated with an exercise's level of difficulty. This observation reaffirms the reality that difficult exercises (e.g., tandem stance with yaw head movements, which reduces the base of support in the medial-lateral direction) challenge balance in the roll direction. Melzer et al. also found that fallers had significantly higher trunk sway in the roll direction than the non-fallers for balance exercises performed with eyes open and eyes closed [55].

Among the statistical descriptors, the trend across six trials was ranked highest in terms of feature importance. This suggests that the PT involved in this study strongly considered changes in performance over all six trials when evaluating participants. Average performance (i.e., mean) and worst performance (i.e., max) among the six trials were the least important statistical descriptors, whereas the best performance (i.e., min) among the six trials was the second most important statistical descriptor. The implication of these results is that one poorly performed trial may not affect the overall rating assigned to the performance, while a well-performed trial may drive the overall rating.

This pilot study has several limitations. First, only one PT provided the ratings that were used as the ground truth, and we did not assess intra-rater reliability. If the intra-rater reliability was low (e.g., two sets that appear to be the same are rated differently), it could have negatively impacted classification accuracy. However, if the intra-rater reliability was low it would affect both the accuracy of the SVM and the participant's self-assessments. Second, we only investigated standing exercises, but standard balance training programs include weight shifting, modified center of gravity, and gait exercises [10], [12], [13]. Different feature extraction methods are likely needed to automate ratings for these additional types of exercises. Third, PT ratings were unevenly distributed across classes. To mitigate this effect, we considered F1 scores when evaluating classification performance. Still, the uneven distribution across classes could suggest why performance was lower in less prevalent classes. Last, the text accompanying the numerical values in the two rating scales used by the participants and PT to assess balance performance differed slightly to reflect the language that each group would use to describe balance performance; the lack of a single scaled used by both the participants and the PT could have affected the outcomes.

Automated balance performance assessments have the potential to augment existing in-home balance training programs. In particular, automated assessments could guide participants in the selection of appropriate exercises, potentially allowing for more flexibility compared to a fixed program, since multiple exercise options exist at any given difficulty level. In addition to providing greater flexibility automated assessment could impact program effectiveness, since training performed at an appropriate level has been shown to lead to greater improvements in balance [12], [56]. Furthermore, automated assessments have the potential to provide balance trainees with immediate feedback following the completion of an exercise, which may increase compliance and motivation, a general problem with traditional in-home training programs [22]. Finally, the accurate identification of difficult/unsafe exercises is likely to reduce the performance of unsafe exercises in the absence of a PT, thereby decreasing the risk of falls in the home [57]. Aside from an increase in the number of losses of balance and an increase in fall risk, exercises that exceed an individual's capabilities typically do not have positive effects on training outcomes [56].

Future work could include studies involving various levels of PT supervision. For example, a non-supervised study performed in the clinic or in the home could leverage a device similar to the smart phone balance trainer described in [13] to assess balance performance by collecting and analyzing trunk-based sway data and step out data (either manually entered by trainees or potentially automatically detected). These assessments could then either be accessed by PTs to form the basis of subsequent balance training regimes or an automated recommender system could be implemented.

## V. CONCLUSION

Using ML techniques, we successfully learned a mapping from trunk sway data from a single IMU to a PT's assessment of balance performance. Compared to self-assessment ratings, the automatically generated ratings more closely agreed with the PT. On a three-level scale, the model achieved an accuracy of 82%. The results of this study could be used to provide in-home balance assessments during unsupervised or semi-supervised balance training programs. Such automated assessments could lead to a reduction in PT consultation time, an increase in compliance and an overall improvement in the effectiveness of in-home balance training programs.

ACKNOWLEDGMENTS

This work was supported by NIH under award No. 1R21DC012410-01A1 and NSF under grant No. IIS-1553146. Any opinions, findings, and conclusions or recommendations solely the responsibility of the authors and does not necessarily represent the official views of the NIH or NSF. We thank Wendy Carender and Catherine Kinnaird for assisting with the design of the rating scales.